\documentclass[12pt]{article}
\usepackage[totalheight = 23cm, totalwidth = 17cm]{geometry}
\usepackage{graphicx, subfigure}
\newcommand{\nin}{\noindent}
\newcommand{\be}{\begin{equation}}
\newcommand{\ee}{\end{equation}}
\newcommand{\bea}{\begin{eqnarray}}
\newcommand{\eea}{\end{eqnarray}}
\newcommand{\br}{\hskip .25cm/\hskip -.25cm}

\newcommand{\nn}{\nonumber\\}

\newcommand{\ol}{\overline}
\usepackage{amsmath}

\begin{document}

\begin{center}
{\Large{\bf Comments on branon dressing and the Standard Model}}

\vspace{1cm}

J. Alexandre\footnote{jean.alexandre@kcl.ac.uk} and D. Yawitch\footnote{darrel.yawitch@kcl.ac.uk}\\
Department of Physics, King's College London, WC2R 2LS, UK

\vspace{2cm}

{\bf Abstract}

\end{center}

\nin
This technical note shows how Electrodynamics and a Yukawa model are dressed after integrating out perturbative brane fluctuations, 
and it is found that first order corrections in the inverse of the brane tension occur for the
fermion and scalar wave functions, the couplings and the masses. Nevertheless, field redefinitions actually
lead to effective actions where only masses are dressed to this first order.
We compare our results with the literature and find discrepancies at the next order, which, however, 
might not be measurable in the valid regime of low-energy brane fluctuations.

\vspace{2cm}

\section{Introduction}

In the context of braneworld scenarios and higher dimensional theories, branons are modes
which correspond to perturbative quantum fluctuations of the brane, about its equilibrium position in 
extra dimensions. Using an  effective description of a four-dimensional brane embedded in a 
five-dimensional universe, branons can be thought of as a scalar field living on the brane,  
representing fluctuations in the position of the brane, as measured in the fifth dimension \cite{branons}. 
Branons can then be thought of as particles which couple to the Standard Model (SM), with a dimensionful coupling 
constant, proportional to the inverse of the brane tension $f^4$.

Branons are interesting to study in the context of Cosmology and Collider Physics, since there could
be phenomenological effects arising from branons, that may be detectable in colliders as the LHC . 
In this paper we study the coupling of branons to Electrodynamics and a Yukawa model, and the corresponding possible
phenomenological effects. We will see that, although the interaction of branons with SM 
particles is suppressed by $f^{-4}$, the derivative interactions between branons and the SM can compensate 
this suppression. The idea is then to integrate out branon degrees of freedom, in order to obtain 
the effective theory for the SM degrees of freedom. In the low energy approximation for branons, 
the coupling to the SM is quadratic in branons, 
such that the integration of the latter can be done exactly. As a consequence, the one-loop result is exact, and
will be expanded in powers of the dimensionful coupling constant $f^{-4}$. 
The resulting effective theory contains corrections 
to SM interactions, but also new interactions, such as four-fermion interactions, all of dimensionality larger than 4. 
We will concentrate here on the corrections to the SM, since the other interactions are suppressed by higher 
orders of the inverse brane tension, $f^{-8}$.

It has been shown in previous studies \cite{branons} that warped extra dimensions lead to massive branons, 
but we will consider a flat extra dimension, leading to massless branons: the dressing effects that we study here are mainly due
to UV dynamics, where the branon mass does not play an important role.

Similar studies were done in \cite{couplingSM}, where the authors look at the coupling of branons to the SM, 
and integrate the former to obtain the corresponding effective action for SM particles. 
We compare this to our approach, and find the same conclusion to the order $f^{-4}$, but 
a discrepancy to the next order $f^{-8}$, for the Yukawa model.
The reason of the discrepancy is that the authors of \cite{couplingSM} assume equations of motion to hold, without taking into 
account the dressing from branons. We do not assume that degrees of freedom of the SM satisfy equations of motion, and 
find that the Yukawa coupling gets a correction. 
Nevertheless, we argue that this $f^{-8}$ correction is certainly negligible in the low-energy approximation, 
where the branon model is valid.

The paper is structured as follows.
Section 2 describes the construction of the low energy theory describing the interaction branons/Electrodynamics. 
Starting from the general expression for the Electrodynamics action in curved space time, we expand the metric in 
terms of the branon degree of freedom,
which leads us to the interactions branon/Electrodynamics. 
We derive then, in section 3, the effective action for Electrodynamics, obtained after exact integration
of branons. We give technical comments on this derivation, by comparing our work to the one given in \cite{couplingSM}. The 
agreement between our approaches, in this specific case, is a consequence of gauge symmetry. Indeed, the gauge coupling and the 
fermion wave function renormalization need to get the same corrections for gauge symmetry to hold, such that, 
after a redefinition of the fields,
the correction to the gauge coupling exactly vanishes, and only the fermion mass gets a correction. Another important fact
is that the Maxwell free theory is conformally invariant, independently of equations of motion, such that the gauge field does not get any correction.
Section 4 shows how a Yukawa model is dressed by branons. In this case, because no gauge invariance is required, we 
find that the Yukawa coupling does get a correction, leading to the discrepancy with \cite{couplingSM}.
Our conclusion questions the concept of branon, which, in order to be well defined, needs a low energy approximation, where
the branon dressing might not be measurable in experiments.

\section{Branon/Electrodynamics interactions}

We consider a 5-dimensional flat Universe with generic coordinates $X^M=(x^\mu,y)$, where $x$ are the coordinates on the brane,
which is defined by the equation $y=Y(x)$. The brane coordinates are denoted with the indices
$\mu,\nu$, and the induced metric $h_{\mu\nu}$ on the brane is 
\be\label{induced}
h_{\mu\nu}(x)=\eta_{\mu\nu}-\partial_\mu Y\partial_\nu Y.
\ee
If $f^4$ is the brane tension, the brane action is then 
\bea
S_{brane}&=&-f^4\int d^4x\sqrt{-h}\\
&=&-f^4\int d^4x\left( 1-\frac{1}{2}\eta^{\mu\nu}\partial_\mu Y\partial_\nu Y-\frac{1}{8}(\eta^{\mu\nu}\partial_\mu Y\partial_\nu Y)^2
+\cdots\right), \nonumber
\eea
where dots represent higher orders in derivatives of $Y$.
Our dynamical variable is the canonically normalized branon field $\phi=f^2Y$, with mass dimension 1, and
the brane ground state is $Y=0$. Ignoring field-independent terms, since we are looking at an effective theory in flat space time,
and taking into account the low-energy branon approximation, the resulting effective action for branons is then
\be\label{Sbranon}
S_{branon}=\int d^4x\left( \frac{\eta^{\mu\nu}}{2}\partial_\mu\phi\partial_\nu\phi\right),
\ee
and describes a free theory. Interactions will occur, though, with particles propagating in the brane over Cosmological distances, 
and we now derive the corresponding action for Electrodynamics.

In what follows, Latin indices refer to the local inertial frame and are contracted with $\eta_{ab}$,
whereas Greek indices are contracted with the induced metric $h_{\mu\nu}$. 
The action describing Electrodynamics in curved background is \cite{birrell}
\bea\label{Sem}
S_{em}&=&\int d^4x\sqrt{-h} 
\Bigg\{ -\frac{1}{4}F_{\mu\nu}F_{\rho\sigma}h^{\mu\rho}h^{\nu\sigma}+
\frac{i}{2}\ol\psi\gamma^a e_a^\mu\left(\partial_\mu+\Gamma_\mu\right)\psi\nn
&&~~~~-\frac{i}{2}e_a^\mu\left(\partial_\mu\ol\psi+\ol\psi\Gamma_\mu\right)\gamma^a\psi
-g_0\ol\psi e^\mu_a\gamma^aA_\mu\psi -m_0\ol\psi\psi\Bigg\},
\eea
where the spin connection is 
\be
\Gamma_\mu=\frac{1}{8}[\gamma^a,\gamma^b]e_a^\nu\nabla_\mu e_{b\nu},
\ee
and the gamma matrices $\gamma^a$ are defined in the local inertial frame, therefore satisfying
\be
\{\gamma^a,\gamma^b\}=2\eta^{ab}.
\ee
Neglecting higher orders in the branon derivatives $\partial Y$, we find the following approximate vierbeins on the brane
\bea
e_\mu^a&=&\delta_\mu^a-\frac{1}{2}\partial_\mu Y\partial^a Y+{\cal O}(\partial Y)^4\nn
e_a^\mu&=&\delta_a^\mu+\frac{1}{2}\partial_a Y\partial^\mu Y+{\cal O}(\partial Y)^4.
\eea
which lead to the expected definition $e_\mu^a e_\nu^b\eta_{ab}=h_{\mu\nu}$, up to
higher orders in $\partial Y$. 
The Christoffel symbols for the induced metric are then
\be
\Gamma^\rho_{~\mu\nu}=-(\partial^\rho Y)(\partial_\mu\partial_\nu Y)+{\cal O}\left( \partial(\partial Y)^4\right) ,
\ee
and the spin connection is
\be\label{spinconnection}
\Gamma_\mu=\frac{1}{8}[\gamma^a,\gamma^b](\partial_a\partial_\mu Y)(\partial_b Y)+{\cal O}\left( \partial(\partial Y)^4\right) ,
\ee
such that the total action for the system branons/Electrodynamics is, after integrations by parts,
\bea\label{Stotal}
S_{total}[\phi,\ol\psi,\psi,A]&=&
\int d^4x\Bigg\{-\frac{1}{4}F^2+\ol\psi(i\br\partial-g_0\br A)\psi-m_0\ol\psi\psi\\
&&+\frac{1}{2}\partial^\mu\phi\partial_\mu\phi+\frac{1}{8f^4}F_{\mu\nu}F_{\rho\sigma}\eta^{\nu\sigma}
\left[\eta^{\mu\rho}(\partial\phi)^2-4\partial^\mu\phi\partial^\rho\phi\right] \nn
&&+\frac{1}{2f^4}\ol\psi\left[ \br\partial\phi\partial^\mu\phi
-(\partial\phi)^2\gamma^\mu\right](i\partial_\mu-g_0A_\mu)\psi\nn
&&+\frac{i}{2f^4}\ol\psi\left(\partial^2\phi\br\partial\phi-\partial^\mu\phi\partial_\mu\br\partial\phi\right)\psi
+\frac{1}{2f^4}(\partial^\mu\phi\partial_\mu\phi)m_0\ol\psi\psi\Bigg\},\nonumber
\eea
where the indices are raised and lowered with the Minkowski metric.
We note that, although the coupling between branons and
Electrodynamics is proportional to $f^{-4}$, it is actually not negligible, because of derivative interactions. 
Finally, the Greek indices appearing in the effective action (\ref{Stotal}) all
denote flat four-dimensional space time coordinates.

\section{Effective action for Electrodynamics}

\subsection{Derivation}

We assume from now on that branons have energies up to some value $\epsilon f$, where $\epsilon<1$ for the low-energy
branon approximation (\ref{Sbranon}) to be valid.
We integrate out their degrees of freedom from the theory described by the action (\ref{Stotal}), which can be done exactly,
since this action is quadratic in the branon field. 
The resulting effective action for Electrodynamics is then
\be\label{Seff}
S_\epsilon[\ol\psi,\psi,A]=\int d^4x \Bigg\{-\frac{1}{4}F^2+\ol\psi(i\br\partial-g_0\br A)\psi-m_0\ol\psi\psi\Bigg\}
+\frac{1}{2}\mbox{Tr}_\epsilon\left\lbrace \ln\left( \frac{\delta^2 S_{total}}{\delta\phi\delta\phi}\right)\right\rbrace  ,
\ee
where the trace is taken over the branon momentum $0\le|p|\le\epsilon f$. This regularization
does not contradict gauge invariance, since the momenta of 
photons and fermions are not restricted, but only the branon momentum is, as can be seen from the Fourier transform
(\ref{fourier}), when the trace is taken ($p+q=0$). The second functional derivative of $S_{total}$ is 
\bea
\frac{\delta^2 S_{total}}{\delta\phi_x\delta\phi_y}&=&-\partial^2\delta^{(4)}(x-y)
-\frac{m_0}{f^4}\partial^\mu\left(\ol\psi\psi\partial_\mu\delta^{(4)}(x-y)\right)\\
&&-\frac{1}{4f^4}\partial_\mu\left(F^2\partial^\mu\delta^{(4)}(x-y)\right)
+\frac{1}{f^4}\partial_\mu\left( F^{\mu\nu}F_{\rho\nu}\partial^\rho\delta^{(4)}(x-y)\right)  \nn
&&+\frac{i}{2f^4}\partial_\rho\left[ \ol\psi\left( -\gamma^\rho\partial^\mu\delta^{(4)}(x-y){\cal D}_\mu
-\br\partial\delta^{(4)}(x-y){\cal D}^\rho-2\partial^\rho\delta^{(4)}(x-y)\br{\cal D}\right)\right]  \psi\nn
&&+\frac{i}{2f^4}\partial_\rho\big[\partial^\rho(\ol\psi\br\partial\delta^{(4)}(x-y)\psi)
-\ol\psi\partial^2\delta^{(4)}(x-y)\gamma^\rho\psi\nn
&&~~~~~~~~~~~~~~~~~~~+\ol\psi\partial^\rho\br\partial\delta^{(4)}(x-y)\psi
-\partial_\nu(\ol\psi\gamma^\nu\partial^\rho\delta^{(4)}(x-y)\psi\big]\nonumber,
\eea
where ${\cal D}_\mu=\partial_\mu+ig_0A_\mu$,
and has the following Fourier transform
\bea\label{fourier}
&&\int d^4xd^4y~\frac{\delta^2 S_{total}}{\delta\phi_x\delta\phi_y}~e^{ipx+iqy}\nn
&=&p^2(2\pi)^4\delta^{(4)}(p+q)-\frac{pq}{f^4}\int_km_0\ol\psi(k)\psi(p+q-k)\nn
&&-\frac{pq}{4f^4}\int_kF^{\mu\nu}(k)F_{\mu\nu}(p+q-k)+\frac{p_\mu q^\rho}{f^4}\int_kF^{\mu\nu}(k)F_{\rho\nu}(p+q-k)\nn
&&-\frac{i}{2f^4}\int_k\int_l\ol\psi(k)\left[(\br p~ q^\mu+\br q~p^\mu){\cal D}_\mu(l)
-2pq\br{\cal D}(l)\right] \psi(p+q-k-l)\nn
&&+\frac{1}{2f^4}\int_k \ol\psi(k)\left[ p^2\br q~+q^2\br p~-pq(\br q~+\br p~)\right] \psi(p+q-k),
\eea
where 
\be
\int_k(\cdots)=\int\frac{d^4k}{(2\pi)^4}(\cdots).
\ee
We expand then the logarithm in the effective action (\ref{Seff}) around the diagonal part 
(proportional to $\delta^{(4)}(p+q)$), and keep the first order in the inverse brane tension, to obtain
\bea\label{trace}
&&\frac{1}{2}\mbox{Tr}_\epsilon\left\lbrace \ln\left( 
\frac{\delta^2 S_{total}}{\delta\phi\delta\phi}\right)\right\rbrace \\
&=&\frac{\epsilon^4}{64\pi^2}\int_km_0\ol\psi(k)\psi(-k) 
-\frac{3\epsilon^4}{256\pi^2}\int_k\int_l\ol\psi(k)i\br{\cal D}(l)\psi(-k-l)\nn
&=&\frac{\epsilon^4}{64\pi^2}\int d^4x\left\{m_0\ol\psi(x)\psi(x)
-\frac{3}{4}\ol\psi(x)i\br{\cal D}(x)\psi(x)\right\rbrace,\nonumber
\eea
where field-independent terms were ignored. 
Note that the contributions for the correction to $F^2$ cancel each other after taking the trace, because
\be
\frac{1}{4}\int_{p,q}\delta^{(4)}(p+q)p^\rho q_\rho F^{\mu\nu}F_{\mu\nu}
=\int_{p,q}\delta^{(4)}(p+q)p_\mu q^\rho F^{\mu\nu}F_{\rho\nu},
\ee
and the terms 
arising from the spin connection (last line in eq.(\ref{fourier})) also cancel in the trace. We will discuss these
two points in the next subsection. We can also note that, although branons are massless, 
no IR divergences appear in the branon loop integrals, because of the derivative interactions compensating the
possible divergences at $p=0$. 
The effective action for Electrodynamics is finally
\be\label{Seffinal}
S_\epsilon[\ol\psi,\psi,A]=\int d^4x \Bigg\{-\frac{1}{4}F^2
+\left(1-\frac{3\epsilon^4}{256\pi^2}\right)\ol\psi(i\br\partial-g_0\br A)\psi
-\left(1-\frac{\epsilon^4}{64\pi^2}\right)m_0\ol\psi\psi\Bigg\},
\ee
and we find a correction to the fermion wave function, the coupling and the mass. Also, as expected, the corrections to
the fermion wave function and to the coupling are the same, which is a consequence of gauge invariance.

\subsection{Comments}

We now explain how the effective action (\ref{Seffinal}) could be obtained differently, as was done in \cite{couplingSM}. \\
In this work, the Authors start from the action
\be\label{LT}
\tilde S=\int d^4x\left\lbrace {\cal L}_{SM}+\frac{1}{2}\partial_\mu\phi\partial^\mu\phi
+\frac{1}{2f^4}\partial_\mu\phi\partial_\nu\phi~T^{\mu\nu}_{SM}\right\rbrace ,
\ee
where ${\cal L}_{SM}$ is the flat space time Standard Model Lagrangian, and $T^{\mu\nu}_{SM}$ is its energy-momentum tensor.
After integration over branons, they find that the first order correction to the Standard Model action is proportional to
\be
\frac{\eta_{\mu\nu}}{f^4}\int d^4x ~T^{\mu\nu}_{SM},
\ee 
which would vanish if the Standard Model was scale invariant and if one assumes the equations of motion to be 
satisfied\footnote{As will be seen in eq.(\ref{tracebis}), the term $F^2$ is conformally invariant, independently of equations of motion, and we repeat here the general argument given in \cite{couplingSM}}.\\
To put our results in a similar context, we bear in mind that  
not only the metric, but also its derivatives depend on the branon field
\be
\partial_\rho h^{\mu\nu}=\frac{1}{f^4}\partial_\rho(\partial^\mu\phi\partial^\nu\phi)+{\cal O}(\partial\phi)^4,
\ee
such that $h^{\mu\nu}$ and $\partial_\rho h^{\mu\nu}$ cannot be considered independent variables.
As a consequence, expanding the following action up to the first order in $(\partial\phi)^2$
\be
S_{em}=\int d^4x\sqrt{-h}\left\lbrace -f^4+{\cal L}_{em}(h^{\mu\nu},\partial_\rho h^{\mu\nu})\right\rbrace,
\ee
we obtain (neglecting constant terms),
\bea\label{Semexpand}
S_{em}&=&\int d^4x\left\lbrace ({\cal L}_{em})_{flat}+\frac{1}{2}\partial^\mu\phi\partial_\mu\phi+
\frac{1}{2f^4}\partial^\mu\phi\partial^\nu\phi(T_{\mu\nu}^{em})_{flat}\right\rbrace \\
&&+\frac{1}{f^4}\int d^4x\left(\frac{\partial{\cal L}_{em}}{\partial(\partial_\rho h^{\mu\nu})}\right)_{flat}
\partial_\rho(\partial^\mu\phi\partial^\nu\phi)+{\cal O}(\partial\phi)^4\nonumber
\eea
where ``flat'' denotes the corresponding quantity for vanishing branon field, and 
\be\label{Tem}
T_{\mu\nu}^{em}=\frac{\delta S_{em}}{\delta h^{\mu\nu}}
=-{\cal L}_{em}~h_{\mu\nu}+2\frac{\partial{\cal L}_{em}}{\partial h^{\mu\nu}}.
\ee
Our results can be understood as follows:
\begin{itemize}

\item The trace of the energy-momentum tensor is
\bea\label{tracebis}
\mbox{tr}\left\lbrace T^{em}_{\mu\nu}\right\}_{flat}
&=&\eta^{\mu\nu}\Bigg\lbrace -\frac{1}{4}F_{\rho\sigma}F_{\omega\tau}\eta^{\rho\omega}
\left(-\eta_{\mu\nu}\eta^{\sigma\tau}+4\delta^\sigma_\mu\delta^\tau_\nu\right)\nn
&&-\frac{i}{2}\eta_{\mu\nu}(\ol\psi\br\partial\psi-\partial_\rho\ol\psi\gamma^\rho\psi)
+\frac{i}{2}(\ol\psi\gamma_\nu\partial_\mu\psi-\partial_\mu\ol\psi\gamma_\nu\psi)\nn
&&-g_0\ol\psi(-\br A\eta_{\mu\nu}+\gamma_\mu A_\nu)\psi+\eta_{\mu\nu}m_0\ol\psi\psi\Bigg\rbrace \nn
&=&-\frac{3}{2}i(\ol\psi\br\partial\psi-\partial_\rho\ol\psi\gamma^\rho\psi)+3g_0\ol\psi\br A\psi+4m_0\ol\psi\psi,
\eea
where the following derivative has been used
\be
\left( \frac{\partial e^\rho_a}{\partial h^{\mu\nu}}\right)_{flat}=\frac{1}{2}\delta_\mu^\rho\eta_{a\nu},
\ee
and leads to the corrections we obtain.
Note that the ratio between the coefficients of the correction to the coupling and the correction to the mass is 3/4, 
which was found in eq.(\ref{Seffinal}). The
Maxwell term $F^2$ doesn't get any correction since its energy-momentum tensor is traceless anyway;

\item The only term in the Lagrangian which contains derivatives of the metric is the spin connection. We have seen that
the corresponding correction vanishes, which is necessary to maintain gauge invariance, since this correction would 
dress the fermion kinetic term and not the coupling. Hence the second line of the expansion (\ref{Semexpand}) should
not play a role, and this can be understood since the corresponding correction is
\bea
&&\mbox{Tr}\left\lbrace \frac{\delta^2}{\delta\phi\delta\phi}
\int d^4x\left(\frac{\partial{\cal L}_{em}}{\partial(\partial_\rho h^{\mu\nu})}\right)_{flat}
\partial_\rho(\partial^\mu\phi\partial^\nu\phi)\right\rbrace \nn
&=&2\int d^4x\int d^4y~\delta^{(4)}(x-y)\partial^\mu\partial_\rho
\left(\frac{\partial{\cal L}_{em}}{\partial(\partial_\rho h^{\mu\nu})}\right)_{flat}\partial^\nu\delta^{(4)}(x-y)\nn
&=&\delta^{(4)}(0)\int d^4x~\partial_\rho\partial^\mu\partial^\nu
\left(\frac{\partial{\cal L}_{em}}{\partial(\partial_\rho h^{\mu\nu})}\right)_{flat},
\eea
which is a surface term.
\end{itemize}

\subsection{Phenomenological implication}\label{phen}

The effective action (\ref{Seffinal}) implies a change in the parameters describing Electrodynamics. 
To see what the change is, we consider the following rescaled fermion field
\be\label{rescale}
\Psi=\psi\sqrt{1-\frac{3\epsilon^4}{256\pi^2}},
\ee
which is our new dynamical variable and has
a canonical kinetic term. The resulting effective action for Electrodynamics becomes
\be\label{Seffrenorm}
S_\epsilon[\ol\Psi,\Psi,A]=
\int d^4x \Bigg\{-\frac{1}{4}F^2+\ol\Psi(i\br\partial-g_0\br A)\Psi-m_\alpha\ol\Psi\Psi\Bigg\},
\ee
where the effective mass is,
\be\label{meff}
m_\alpha=\frac{1-\epsilon^4/64\pi^2}{1-3\epsilon^4/256\pi^2}~m_0
\simeq(1-\alpha)m_0,~~~~\mbox{with}~~~~\alpha=\frac{\epsilon^4}{256\pi^2}<<1.
\ee
As a consequence, after field redefinition, we find that only the fermion mass
gets a correction from branon dressing, with a value agreeing with \cite{couplingSM}, since $\alpha<<1$, 
and which leads to hardly no measurable effect.

It is worth remembering that in experiments we measure only the dressed parameters. Therefore, in order to distinguish the branon-dressed parameters from the bare ones, 
we would need to compare an experiment with branons with a similar experiment without branons, which is difficult to set up.

\section{Coupling to a Yukawa model}

In the previous example of Electrodynamics, we found no dressing for the coupling after field redefinition, since
corrections to the coupling, before field redefinition, have to be the same as corrections to the fermion kinetic term, 
because of gauge invariance.
We consider here a Yukawa model, not ``protected'' by a gauge symmetry, and we will see that the coupling does get a 
correction, to the order $f^{-8}$ though.

\subsection{Corrections to the bare Lagrangian}

We briefly present here similar arguments for branons coupled to a Yukawa model.\\
We start with the action
\bea\label{SY}
S^Y&=&\int d^4x\sqrt{-h}\Bigg\lbrace\frac{h^{\mu\nu}}{2}\partial_\mu\xi\partial_\nu\xi
+\frac{i}{2}e^\mu_a(\ol\psi\gamma^a\partial_\mu\psi-\partial_\mu\ol\psi\gamma^a\psi+\ol\psi[\gamma^a,\Gamma_\mu]\psi)\nn
&&~~~~~~~~~~~~~~~~~~~~~~~-f^4-\lambda_0\xi\ol\psi\psi-\frac{M_0^2}{2}\xi^2-m_0\ol\psi\psi\Bigg\rbrace ,
\eea
where the induced metric $h_{\mu\nu}$ is given by eq.(\ref{induced}).
We use here directly the approach given in \cite{couplingSM}, and write the action (\ref{SY}) in the form 
(neglecting higher orders in branon derivatives)
\be
S^Y=\int d^4x\left\lbrace {\cal L}_Y+\frac{1}{2}\partial^\mu\phi\partial^\nu\phi\left(\eta_{\mu\nu}+\frac{1}{f^4}T^Y_{\mu\nu}\right) \right\rbrace,
\ee
where ${\cal L}_Y$ is the Yukawa Lagrangian in flat space time, and $T^Y_{\mu\nu}$ the corresponding energy momentum tensor.
The second functional derivative of $S^Y$ with respect to the branon field has Fourier transform
\be
\frac{\delta^2S^Y}{\delta\phi\delta\phi}=p^2(2\pi)^4\delta^{(4)}(p+q)-\frac{p^\mu q^\nu}{f^4}\tilde T^Y_{\mu\nu}(p+q),
\ee
where $\tilde T^Y_{\mu\nu}$ is the Fourier transform of $T^Y_{\mu\nu}$.
The integration over branons, up to the energy $\epsilon f$, leads to the following effective action (ignoring field independent terms)
\be\label{traceYukawa}
S^Y_\epsilon=S^Y+\frac{1}{2}\mbox{Tr}_\epsilon\left\lbrace \ln\left((2\pi)^4\delta^{(4)}(p+q)
-\frac{p^\mu q^\nu}{p^2f^4}\tilde T^Y_{\mu\nu}(p+q)\right) \right\rbrace.
\ee
When expanding the logarithm, we obtain (again, ignoring field independent terms)
\bea\label{traceYukawaexpand}
S^Y_\epsilon&=&S^Y-\frac{1}{2f^4}\mbox{Tr}_\epsilon\left\lbrace\frac{p^\mu q^\nu}{p^2}\tilde T^Y_{\mu\nu}(p+q)\right\rbrace
+{\cal O}(T^Y)^2\nn
&=&S^Y+\frac{1}{2f^4}\int d^4x ~\eta^{\mu\nu}T^Y_{\mu\nu}(x)+{\cal O}(T^Y)^2
\eea
and the trace over Lorentz indices of $T^Y_{\mu\nu}$ is
\bea\label{TY}
\eta^{\mu\nu}T_{\mu\nu}^Y&=&
-(\partial\xi)^2-\frac{3i}{2}(\ol\psi\br\partial\psi-\partial_\mu\ol\psi\gamma^\mu\psi)\nn
&&+4\lambda_0\xi\ol\psi\psi+2M_0^2\xi^2+4m_0\ol\psi\psi.
\eea
In the situation where $m_0=M_0=0$ (scale invariant theory), the trace (\ref{TY}) can also be written 
\bea\label{TYmassless}
(\eta^{\mu\nu}T_{\mu\nu}^Y)^{massless}&=&\partial_\mu\left( -\xi\partial^\mu\xi+\frac{3i}{2}\ol\psi\gamma^\mu\psi\right) \nn
&&+\xi\left( \partial^2\xi+\lambda_0\ol\psi\psi\right) -3\ol\psi\left( i\br\partial\psi-\lambda_0\xi\psi\right),
\eea 
and is therefore the sum of a divergence plus terms which vanish if one assumes the equations of motions to hold, {\it before dressing}. 
But because of branon dressing, these equations of motion hold only up to terms of order $f^{-4}$, such that, because of the overall
additional factor $f^{-4}$ in eq.(\ref{traceYukawaexpand}), we expect a difference of order $f^{-8}$ with \cite{couplingSM}. Therefore 
we do not assume any cancellation of the trace (\ref{TYmassless}), and, from eq.(\ref{traceYukawaexpand}) and by analogy with the calculations 
performed for the Electrodynamics case, we are led to the following effective Lagrangian
\be\label{LeffYukawa}
{\cal L}_{eff}=\frac{1}{2}(1-2\alpha)(\partial\xi)^2+i(1-3\alpha)\ol\psi\br\partial\psi
-(1-4\alpha)\left( \lambda_0\xi\ol\psi\psi+\frac{M_0^2}{2}\xi^2+m_0\ol\psi\psi\right) ,
\ee
where $\alpha$ is given is eq.(\ref{meff}).

\subsection{Phenomenological implications}

In terms of the rescaled fields
\be
\Xi=\xi\sqrt{1-2\alpha}~~~~~~~~\mbox{and}~~~~~~~\Psi=\psi\sqrt{1-3\alpha},
\ee
we obtain from the effective Lagrangian (\ref{LeffYukawa})
\be
{\cal L}_{eff}=\frac{1}{2}(\partial\Xi)^2+i\ol\Psi\br\partial\Psi-\lambda_\alpha\Xi\ol\Psi\Psi 
-\frac{M^2_\alpha}{2}\Xi^2-m_\alpha\ol\Psi\Psi,
\ee
where the effective masses are
\bea
M_\alpha&=&M_0\sqrt\frac{1-4\alpha}{1-2\alpha}=\left(1-\alpha-\frac{5}{2}\alpha^2+{\cal O}(\alpha^3)\right)M_0\nn
m_\alpha&=&m_0\frac{1-4\alpha}{1-3\alpha}=\left(1-\alpha-3\alpha^2+{\cal O}(\alpha^3)\right)m_0,
\eea
and the effective coupling is
\be\label{lambdalpha}
\lambda_\alpha=\frac{\lambda_0~(1-4\alpha)}{\sqrt{1-2\alpha}(1-3\alpha)}=\lambda_0\left(1-\frac{5}{2}\alpha^2 
+{\cal O}(\alpha^3)\right).
\ee
Hence, although the correction (\ref{lambdalpha}) occurs at the order $f^{-8}$ only, 
it does not vanish exactly, as one could conclude from \cite{couplingSM}.
Note that this $f^{-8}$ correction has nothing to do with higher orders
in the expansion of the logarithm in eq.(\ref{traceYukawa}), 
since these higher orders lead to four-fermion interactions and other operators 
with dimension larger than 4. The Yukawa coupling gets corrections from the first order only, in the expansion of the logarithm
appearing in eq.(\ref{traceYukawa}).\\
Nevertheless, if one considers a low-energy branon regime, consistent with the initial approximation (\ref{Sbranon}), one should
assume that $\epsilon<1$, such that the correction (\ref{lambdalpha}) satisfies
\be\label{deltalambda}
\frac{5}{2}\alpha^2<4\times10^{-7},
\ee
and the discrepancy with \cite{couplingSM} actually hasn't got measurable consequences, since
the Yukawa coupling of the SM is measured via fermion masses: the correction (\ref{deltalambda}) is smaller than the error bars 
on the Cabibbo-Kobayashi-Maskawa matrix.

Again, as with the electrodynamics case it is worth remembering that in experiments we measure only the dressed parameters and the comments at the end of section \ref{phen} apply here as well.

\section{Conclusion}

We discussed in this paper, in the framework of brane models,
how two sectors of the Standard Model are dressed by brane fluctuations. 
This was done by integrating out a massless scalar degree of freedom, associated to perturbative brane fluctuations, the branon.
In order to check results previously published, we performed this integration explicitly, in the case of Electrodynamics, without assuming any symmetry 
of the energy-momentum tensor, and find the expected result. But in the situation of the Yukawa interaction, we find that 
the coupling does get a correction, which
differs from previous studies, but with a difference hardly measurable experimentally.

More generally, the concept of branon gives an elegant effective description of low energy brane fluctuations, but the 
feedback on the coupling constants of the Standard Model is of order $f^{-8}$ only,
making this effective description hard to test experimentally, especially if the brane tension is expected to be 
larger than (100 GeV)$^4$ \cite{creminelli}.
Also, it is not really clear how to interpret the cut off $\epsilon f$ in branon energy, making even more difficult the 
experimental tests. One possibility is to assume that $\epsilon f$ corresponds to a centre of mass energy in a collision,
in which case the branon dressing can be compared to quantum corrections. But the corresponding dressing of
coupling constants, of order $f^{-8}$, can be in the error bars of the experimental measurements if the brane tension is large.
Bounds have been discussed in \cite{couplingSM} though.
Another possibility is to assume a branon bath of typical temperature $\epsilon f$, in a finite temperature collision, 
but it is not clear how efficiently branons thermalize to Standard Model particles.
One can also assume that branons are produced by an astrophysical catastrophic event, but in this case, the 
parameter $\epsilon$ is then certainly so small, that no effect can be detected.

Furthermore, we note that, in order to distinguish the branon-dressed parameters from the bare ones, 
we would need to compare an experiment with branons with a similar experiment without branons, which is difficult to set up.
It seems more realistic to check the effects of branons at the next order in the inverse brane tension, 
since new interactions appear when the logarithm in the effective action (\ref{Seff}) is developed further. 
These effects are of order $f^{-8}$ though, and have been calculated in \cite{couplingSM}. 
The discrepancy we find here with this reference would not change their 
results substantially, since the corresponding correction occurs at the order $f^{-16}$.

We also would like to remark that the initial approximation (\ref{Sbranon}) takes into account the 
first order only in $(\partial\phi)^2$, to get the action describing the interaction branon-Standard Model.
This approximation is based on the occurrence of low energy branons, and contains two derivatives of the branon field.
But the resulting coupling $F^2(\partial\phi)^2$ contains four derivatives, such that, should we follow a strict
gradient expansion scheme, we would have to take into account also terms of order $(\partial\phi)^4$ in the expansion of 
the induced metric $h_{\mu\nu}$. This was not done, and would spoil the exact integration over branon degrees of freedom. 
These additional terms, though, might be relevant in the limit of the parameter $\epsilon$ goes to 1, where the low-energy
branon approximation is not valid anymore. These ambiguities show that a proper study of brane fluctuations feedback on the 
Standard Model might need to be non-perturbative. The concept of branon degree of freedom would then be difficult to define, 
but phenomenological effects might be more realistically predictable.

\section*{Acknowledgements}

We would like to thank Malcolm Fairbairn for useful discussions. The work of
D. Y. is supported by the STFC, UK, and this work is also partly supported by the Royal Society, UK.

\end{document}